\documentclass[11pt]{revtex4-1}

\usepackage{graphicx}

\begin{document}

\title{A third alternative to explain recent observations: Future deceleration}

\author{Subenoy Chakraborty\footnote {schakraborty@math.jdvu.ac.in}}
\affiliation{Department of Mathematics, Jadavpur University, Kolkata 700032, West Bengal, India.}

\author{Supriya Pan\footnote {span@research.jdvu.ac.in}}
\affiliation{Department of Mathematics, Jadavpur University, Kolkata 700032, West Bengal, India.}

\author{Subhajit Saha\footnote {subhajit1729@gmail.com}}
\affiliation{Department of Mathematics, Jadavpur University, Kolkata 700032, West Bengal, India.}

\begin{abstract}
In the present work we discuss  a third alternative to explain the latest observational data concerning the accelerating Universe and its different stages. The particle creation mechanism in the framework of non-equilibrium thermodynamics is considered as a basic cosmic mechanism acting on the flat FRW geometry. By assuming that the gravitationally induced particle production occurs under ``adiabatic" conditions, the deceleration parameter is expressed in terms of the particle creation rate which is chosen as a truncated power series  of the Hubble parameter. The model shows the evolution of the Universe starting from inflation to the present late time acceleration and it also predicts future decelerating stage.\\\\
Keywords: Particle creation, Isentropic process, Future deceleration.\\
PACS Numbers: 98.80.Hw, 04.40.Nr, 05.70.Ln, 95.30.Tg\\\\

\end{abstract}

\maketitle

There are two known distinct approaches to explain the recent accelerated expansion of the Universe as predicted by supernovae Ia and complementary  observations \cite{Riess1, Spergel1, Tegmark1}. Within the framework of Einstein gravity, an unknown type of matter component (dark energy) was introduced having a large negative pressure to explain this accelerating phase. Secondly, the Einstein's gravity theory was modified and the extra terms in the geometric part are interpreted as hypothetical matter component to explain the observational predictions. In both the approaches, attempts have been made to explain only the present accelerated expansion, and, as such, there is no concern about the past or future evolution of the Universe. In the present work, we make an attempt to explain not only the present accelerated expansion but also the past evolution of the Universe starting from a primeval  inflation epoch and further make a prediction about the future evolution of the Universe. The approach adopted here is based on  the mechanism of particle creation in the framework of non-equilibrium thermodynamics \cite{PRI, CLW92} (for an associated  kinetic description see \cite{LB2014}).

The Universe is assumed to be well described by a flat Friedmann-Robertson-Walker (FRW) geometry in agreement with inflation and the cosmic microwave background (CMB) observations. Due to the  gravitationally induced  particle creation  mechanism,  the Universe evolves  like  an open thermodynamical system where the number of fluid particles is not conserved ($N^{\mu}_{;\mu}\neq 0$) \cite{PRI, CLW92, LB2014} (for an earlier effective description of particle production in terms of bulk viscosity see \cite{Zel'dovich1, Barrow}). In the present context, the particle flux satisfies the balance equation:

\begin{equation}
N^{\mu}_{;\mu} \equiv \dot{n}+\Theta n=n\Gamma,
\end{equation}
where $\Gamma$ stands for the rate of change of the number of particles ($N=na^3$) in a comoving volume $a^3$ ($a$ is the scale factor of the FRW model) and $\Theta$ is the fluid expansion scalar. Clearly, positivity of $\Gamma$ indicates the creation of particles while $\Gamma <0$ stands for particle annihilation. It is to be noted that a non vanishing $\Gamma$ is dynamically equivalent to an effective bulk pressure \cite{CLW92, LB2014, Zel'dovich1, Barrow, Hu1, Barrow2, Barrow3, Barrow4} working on the fluid, and, as such,  one can use the methods and techniques of non-equilibrium thermodynamics. However, as discussed long ago by Lima and Germano \cite{LG92}, such scalar processes (bulk viscosity and matter creation) are not equivalent from a thermodynamic viewpoint. The previous  statement about the dynamic behavior can simply be demonstrated in the case of ``adiabatic" particle production as follows \cite{CLW92, LG92}.

From Gibb's equation one may write:

\begin{equation}
Tds=d \left(\frac{\rho}{n}\right)+pd \left(\frac{1}{n}\right),
\end{equation}
and using the balance equation (1), one may write

\begin{equation}
nT\dot{s}=-\Pi \Theta -\Gamma (p+\rho),
\end{equation}
where $T$ is the fluid temperature and $s$ is the specific entropy (per particle). Note that in the above expression, the energy conservation law for an imperfect relativistic simple fluid endowed with creation pressure ($\Pi$) has been considered.  Now, by assuming that creation happens  under ``adiabatic" conditions (see, for instance, \cite{CLW92, Barrow4}), the specific entropy (per particle) does not change, {\it i.e.,} $\dot{s}=0$, and from  Eq. (3) one obtains

\begin{equation}
\Pi=-\frac{\Gamma}{\Theta}(p+\rho).
\end{equation}

Hence the cosmic substratum is not described by a conventional perfect fluid, rather it behaves like an imperfect fluid endowed with a negative pressure describing the time varying comoving number of particles. In other words, although $\dot{s}=0$, still there is entropy production due to enlargement of the phase space of the system ($S \propto N$, where S is the entropy in the comoving volume). Now, by eliminating the creation pressure $\Pi$ from the Einstein's equations ($\kappa =8\pi G$)

\begin{equation}
3H^2=\kappa \rho~,~~\dot{H}=-\frac{\kappa}{2}(p+\rho+\Pi),
\end{equation}
and using (4), it is readily seen that the deceleration parameter reads:

\begin{equation}
q\equiv -\left(1+\frac{\dot{H}}{H^2}\right)= -1+\frac{3\gamma}{2}\left(1-\frac{\Gamma}{\Theta}\right),
\end{equation}
where $\gamma$ is the adiabatic parameter appearing in the equation of state, $p=(\gamma-1)\rho$.

In the present model, the cosmic history is characterized by the fundamental physical quantities namely the expansion rate $H$ and the energy density which can define in a natural way a gravitational creation rate $\Gamma$. From a thermodynamical point of view, $\Gamma$ should be greater than $H$ in the very early universe to consider the created radiation as a thermalized heat bath. So the simplest choice of $\Gamma$ should be $\Gamma\propto H^2$ \cite{AL1} ({\it i.e.,} $\Gamma\propto \rho$) at the very early epoch. The corresponding cosmological solution \cite{LG92, LBC2012, Zimdahl1} shows a smooth transition from inflationary stage to radiation phase and for this adiabatic production of relativistic particles, the energy density scales as $\rho_r \sim T^4$ (black body radiation, for details see Ref. \cite{LBC2012}). 

Recently, it has also been shown \cite{SSS1} that $\Gamma \propto H$ and $\Gamma \propto \frac{1}{H}$ describe respectively the intermediate matter dominated era starting from radiation and the transition from matter dominated era to late time acceleration. Further, Ref. \cite{Chakraborty1} shows that $\Gamma =\Gamma _0, a~constant$ describes the emergent scenario. So a natural question arises: Can

\begin{equation}
\Gamma= \Gamma _0+l H^2+ m H+\frac{n}{H},
\end{equation}

a linear combination of the above four choices describe the whole evolution of the universe? From Eq. (6), the expression for the deceleration parameter implies that there is a transition from deceleration to acceleration or vice-versa at the values of the Hubble parameter given by the cubic equation

\begin{equation}
l H^3+ (m-3\delta) H^2+ \Gamma _0 H+ n= 0,
\end{equation} 
with $\delta= \left(1-\frac{2}{3\gamma}\right)$.

There will be three positive roots of this cubic equation provided\\
$$l> 0, m< 3\delta, \Gamma _0 >0, n< 0,$$ or
\begin{equation}
l< 0, m> 3\delta, \Gamma _0 <0, n> 0.
\end{equation}

In figure 1, we display the variation of $q$ with the Hubble parameter and the three roots correspond to transitions (Inflation $\rightarrow$ Radiation), (Matter dominated era $\rightarrow$ Late time acceleration) and (Late time acceleration $\rightarrow$ Future deceleration). 

\begin{figure}
\includegraphics[height=3in, width=3in]{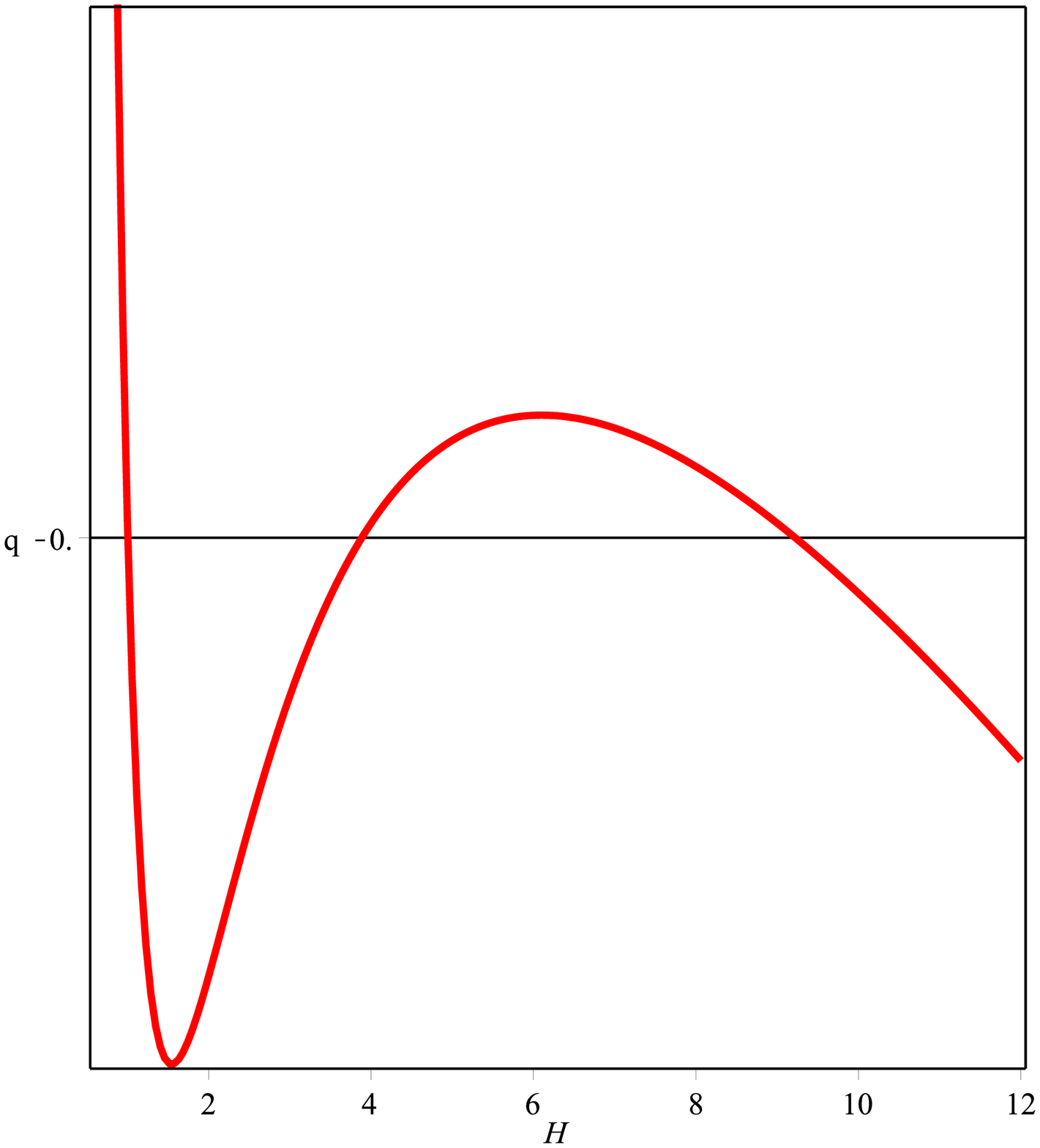}\\
Figure 1 shows the possible future deceleration of the universe. The parameters are $\Gamma_0= 0.1$, $\gamma= 4/3$, $l\approx 0.002$, $m\approx1.47$ and $n\approx -0.07$.
\end{figure}

Further, using the above choice of $\Gamma$, the Hubble parameter can be solved from the Einstein equation as 
\begin{equation}
\frac{ln(H-H_1)}{(H_1-H_2)(H_1-H_3)}+ \frac{ln(H-H_2)}{(H_2-H_1)(H_2-H_3)}+ \frac{ln(H-H_3)}{(H_3-H_1)(H_3-H_2)}= \frac{\gamma a}{2}(t-t_0),
\end{equation}
where $H_1$, $H_2$ and $H_3$ are the positive roots of the cubic Eq. (8). Due to complicated form, we cannot proceed further to solve for the scale factor.\\

In order to show our claim to be true and also due to mathemtical complexity, we start with the simple choice 

\begin{equation}
\Gamma= \Gamma_0+ m H+\frac{n}{H},
\end{equation}

i.e., we put $l= 0$ in the original choice. Now inserting  Eq. (11) into Eq. (4), one can integrate the second Friedmann equation (5) to obtain

\begin{equation}
\frac{|H-H_\alpha|^{H_\alpha}}{|H-H_\beta|^{H_\beta}}= \mu (1+ z)^{-\delta},~~~~\delta= \frac{\gamma}{2} (m-3) (H_\alpha- H_\beta),
\end{equation}

where $\mu$ is an integration constant, $H_\alpha$, $H_\beta$ are the roots of the quadratic equation

\begin{equation}
(m-3) H^2+ \Gamma_0 H+ n= 0,
\end{equation}
i.e., $$H_{\alpha, \beta}= \frac{-\Gamma_0\pm \sqrt{\Gamma_0 ^2- 4 n (m-3)}}{2(m-3)}.$$

Hence for positivity of $H_\alpha$, $H_\beta$ we must have

$$m< 3, \Gamma_0> 0, n< 0~~and~~\Gamma_0^2> 4|n| |m-3|,$$

or

\begin{equation}
m> 3, \Gamma_0< 0, n> 0~~and~~\Gamma_0^2< 4|n| |m-3|.
\end{equation}

Combining with Eq. (9) , we have $m< min(3, 3 \delta)$ or $m> max(3, 3 \delta)$ and other conditions remain same.\\

\begin{figure}
\includegraphics[height=3in, width=3in]{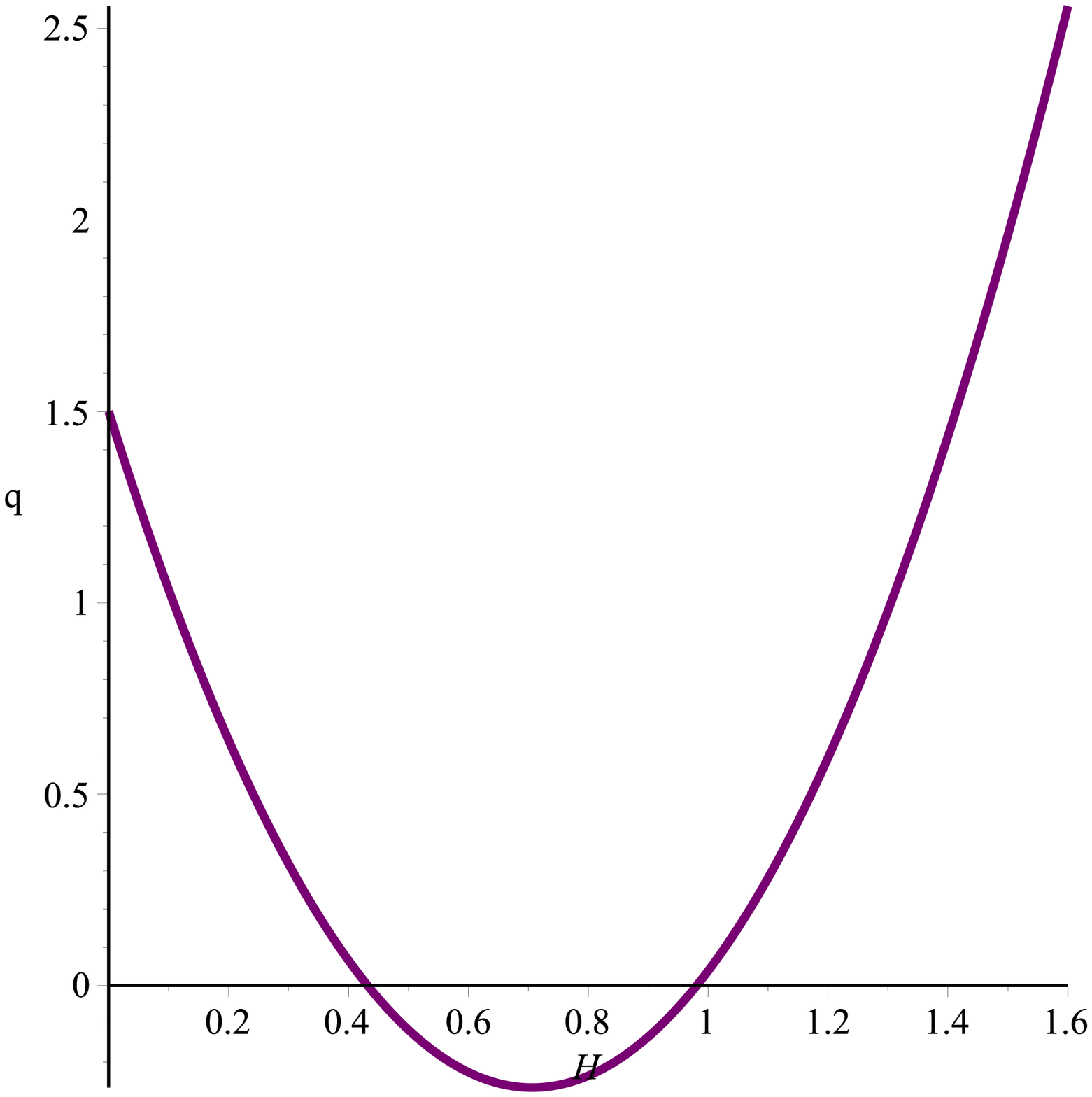}\\
Figure 2 shows the possible future deceleration of the universe for our second choice of the particle creation rate (see Eq. (11)). The parameters are $\Gamma_0= -5$, $m= 5$, $n= 1.5$ and $\gamma= 1.3$.
\end{figure}

Figure 2 shows the variation of $q$ for this choice of $\Gamma$, and from the figure, we see that there are two transitions, ``deceleration $\rightarrow$ late time acceleration" and then again a transition from ``acceleration $\rightarrow$ future deceleration". Also, from this figure, we obtain the values of H at the transition points. Then using Eq. (12) we obtain the corresponding values of $z$ and they turn out to be $z= 0.29$ and $z= -0.34$ respectively. Hence the late time acceleration has started from recent past and there will be again deceleration in future at $z= -0.34$. Further, it is known in the literature \cite{Zimdahl1, SSS1, Lima96} that $\Gamma \propto H^2$ describes the early inflationary scenario.\\

Thus the present phenomenological choice of the particle creation rate,  $\Gamma$ = $\Gamma _0+ l H^2+ m H+ \frac{n}{H}$,  describes the evolution of the universe from inflationary scenario to the present late time acceleration, showing transitions in early era to radiation phase and then again a transition at the end of matter dominated era. Moreover, our  model also predicts a future transition from the present accelerating stage to a future decelerating phase. In this connection, it should be mentioned that similar transient acceleration was considered earlier by Lima and collaborators in the context of scalar-field dominated cosmology \cite{Carvalho1}, as well as, by investigating the cosmic expansion through a kinematic (cosmographic) approach based on Supernovae Ia data \cite{Guimaraes1}.\\

Further, to have some correspondence of our theoretical results with the available observational data, we write the scale factor in power series about the present time (to describe the late time cosmic expansion) as \cite{Guimaraes1}

\begin{equation}
a(t)= 1+ H_0 (t-t_0)-\frac{1}{2!} q_0 H_0 ^2 (t-t_0)^2+ \frac{1}{3!} j_0 H_0 ^3 (t-t_0)^3+\frac{1}{4!} s_0  H_0^4 (t-t_0)^4+ O[(t-t_0)^5],
\end{equation}
where $H_0$, $q_0$, $j_0$ and $s_0$ are the values of the Hubble, deceleration, jerk and snap parameters at the present epoch $t_0$. Then the deceleration parameter can be expressed as

\begin{equation}
q(z)= q_0+ (-q_0-2 q_0^2+ j_0)z+\frac{1}{2} (2q_0+ 8q_0^2+ 8q_0^3-7q_0j_0-4j_0-s_0)z^2+ O(z^3).
\end{equation}

\begin{figure}
\includegraphics[height=3in, width=3in]{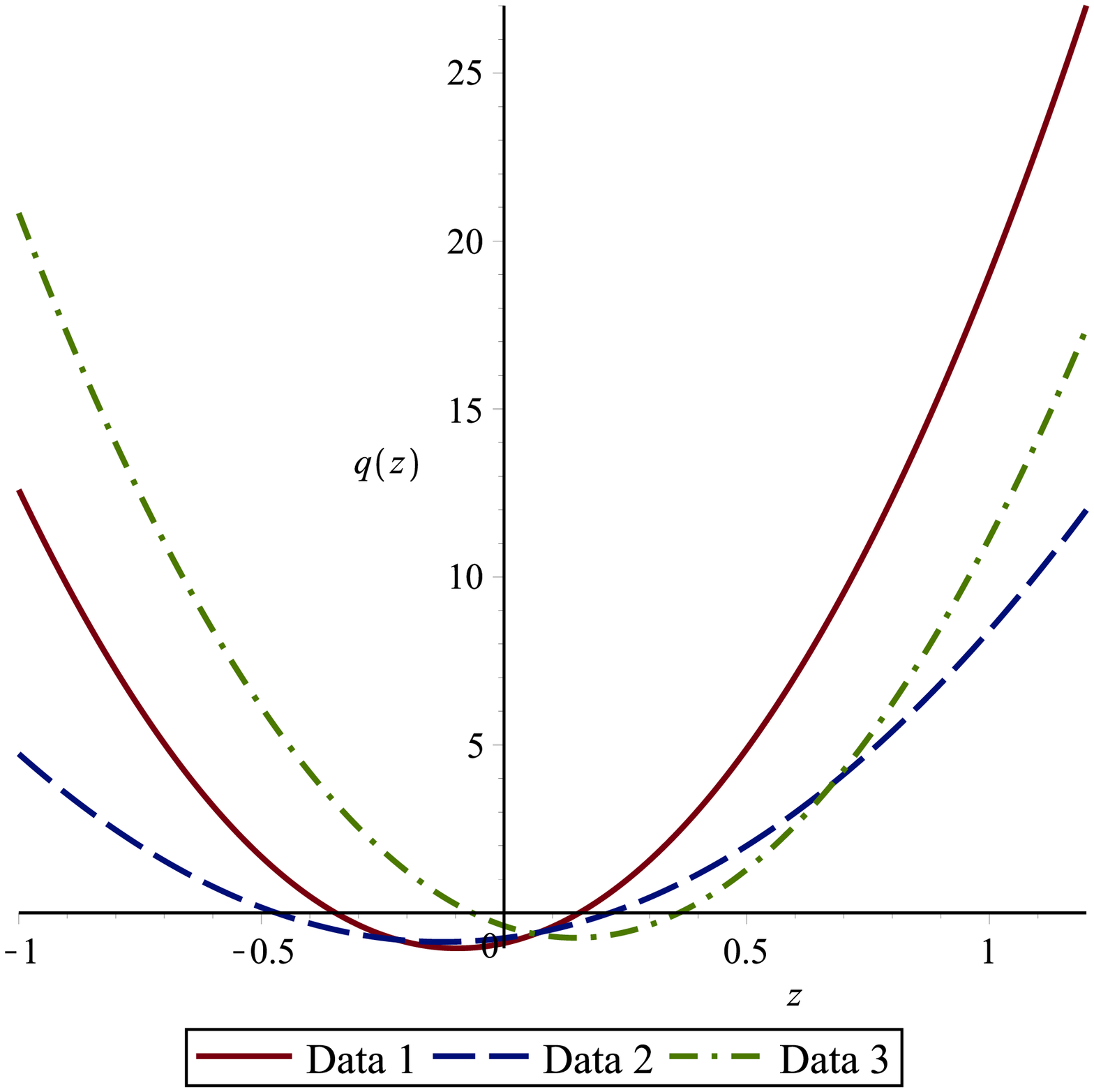}\\
Figure 3 shows the variation of $q$ over $z$ for three observed data sets.
\end{figure}

Note that, if we truncate the power series for the scale factor at fourth order, then $q(z)$ will be quadratic, i.e., $q(z)= q_0+ q_1 z+ \frac{1}{2} q_2 z^2$. We shall now use the cosmographic parameters ($q_0, j_0, s_0$) for three data sets, namely, 192 SNe Ia+ GRBs with CPL parameterization \cite{Wang1} (data 1) and Linear parameterization \cite{Wang1} (data 2) respectively and Supernovae Union 2+ BAO+ OHD+ GRBs \cite{Xu1} (data 3) to draw the graphs of $q$ against the redshift parameter $z$ for Eq. (16).\\

Figure 3 shows the values of $z$ at the transition points for the data sets 1, 2 and 3 which are found to be very close to our $z$ values calculated from Figure 2. Also, all these observed data sets as well as our theoretical predictions are well within the range of the late time acceleration as predicted by the cosmic observations based on Supernovae type Ia \cite{Riess1, Amanullah1}.\\

Summarizing, a model with gravitationally induced  particle production as described here is able to determine  the whole evolution of the Universe starting from an early inflationary era to the present late time acceleration and also predicts a possible transition for a decelerating stage in the future.\\


\begin{acknowledgments}

S. C. is thankful to UGC-DRS programme at the Department of Mathematics. S. P. and S. S. acknowledge, respectively, the Council of Scientific and Industrial Research (Indian Govt.) and UGC-BSR Programme of Jadavpur University for providing research fellowships. The authors are also grateful to J. A. S. Lima for helpful discussions. 

\end{acknowledgments}


\end{document}